\documentstyle[aps,prd]{revtex}
\renewcommand{\baselinestretch}{1.2}
\begin{document}
\large
\title{Relativity in binary systems as root of quantum mechanics and space-time} 
\author{W. Smilga}
\address{Isardamm 135 d, D-82538 Geretsried, Germany}
\address{e-mail: wsmilga@compuserve.com}
\maketitle 
\renewcommand{\baselinestretch}{1.2}

\begin{abstract}
Inspired by Bohr's dictum that ``physical phenomena are observed 
\begin{em}relative\end{em} to different experimental setups", 
this article investigates the notion of \begin{em}relativity\end{em} 
in Bohr's sense, starting from a set of binary elements.
The most general form of information coding within such sets 
requires a description by four-component states. 
By using Bohr's dictum as a guideline a quantum mechanical description
of the set is obtained in the form of a SO(3,2) based spin network. 

For large (macroscopic) sub-networks a flat-space approximation of SO(3,2)
leads to a Poincar\'{e} symmetrical Hilbert space. 
The concept of a position of four-component spinors 
\begin{em}relative\end{em} to macroscopic sub-networks then delivers the 
description of `free' massive spin-1/2 particles with a Poincar\'{e} 
symmetrical Hilbert space.

Hence Minkowskian space-time, equipped with spin-1/2 particles, is obtained 
as an inherent property of a system of binary elements when individual 
elements are described \begin{em}relative\end{em} to macroscopic sub-systems.

\end{abstract}


\renewcommand{\baselinestretch}{1.0}

\section{Introduction}

Spin networks were introduced by R.~Penrose \cite{rp} more than 30 years ago, 
in an attempt to describe the geometry of space-time in a purely combinatorial 
way. 
Penrose studied systems of two-component spinors which represent the simplest 
quantum mechanical objects. 
He was able to show that large systems of such spinors, generate properties of 
angular directions in three-dimensional space. 
Despite this success, the concept of SU(2) based spin networks was considered 
not rich enough to describe also distances \cite{rpwr}. 
Penrose's interest, therefore, turned to more complex twistor objects.
Nowadays spin networks are discussed as models of space-time at Planck scales
\cite{jcb}.

Another attempt to derive space-time from ``binary alternatives" was 
undertaken by C. F. von Weizs\"acker \cite{cfw1,cfw2} and co-workers. 
In this concept a physical particle is considered as an aggregate of
$10^{40}$ binary alternatives, representing the knowledge about the particle.
Although indications for a Poincar\'{e} symmetric space-time structure were 
obtained, the work remained at an experimental stage without having 
established a definite link to empirical particle physics.

The present article describes a new approach to derive space-time as a 
property of general binary systems. 
This goal will be achieved at `normal' (non-Planck) scales. 
As a valuable guide we will repeatedly make use of Bohr's dictum that 
``physical phenomena are observed \begin{em}relative\end{em} to different 
experimental setups" \cite{mj,dmct}. 
Bohr's dictum does not describe any law of nature.
It rather, in a short and precise way, explains how we do physics.
Because of its importance for the understanding of physics, especially at the 
very basic level, in this paper we will speak of \begin{em}Bohr's principle of 
relativity\end{em} or, in short, of \begin{em}Bohr's principle\end{em}.

Bohr's principle has played an important role in the interpretation of 
quantum mechanics. 
It has been used there in the sense that, the description of a quantum 
mechanical phenomenon must include the experimental setup by which
the phenomenon is observed.
Depending on the setup an electron may appear either as a particle or as a 
wave.
We will use Bohr's principle in the sense that 
``particle physics is a discipline that describes \begin{em}microscopic 
phenomena relative to macroscopic setups\end{em}". 

Our approach will start from a set of binary elements, and we will first 
find a way to describe the information contained in a binary element
\begin{em}relative\end{em} to others.
We will see that in the most general case this is achieved by  
four-component states. 
Inspired again by Bohr's principle, in a next step we will introduce a quantum
mechanical description of the binary elements.
This leads to a spin network, made up of Dirac spinors.

Obviously, the Hilbert space of Dirac spinors is by far richer that the state 
space of two-component spinors that was used by Penrose. 
In fact, within this Hilbert space Dirac matrices generate transformations 
that act transitive on the state space. It is well-known that Dirac matrices 
thereby define an irreducible representation of the de~Sitter group SO(3,2).  
A subgroup of SO(3,2) is the (homogeneous) Lorentz group SO(3,1).

In large systems of Dirac spinors we can construct states, where 
some quantum numbers add up to large values as compared to other quantum 
numbers or expectation values.
This then allows for a flat-space approximation by the well-known method of
group contraction \cite{iw,fg}, which approximates the original SO(3,2) 
symmetry by a Poincar\'{e} symmetry. 
The translations of the Poincar\'{e} group then generate a tangential 
Minkowskian space-time continuum.

This Poincar\'{e} invariant basis delivers the framework for a more detailed
description of individual spinors relative to the spin network, by quantum 
mechanical states that carry information about their 
\begin{em}position\end{em} within the network. 
These states satisfy a Dirac equation with finite mass.

Thereby we obtain a quantum mechanical system of identical, massive,
non-interacting spin-1/2 particles defined in a space-time continuum with 
three spacelike and one timelike dimension. This system is invariant under
transformations of the Poincar\'{e} group.

Since the Poincar\'{e} symmetry is only an approximation, we need to examine
possible contributions of higher order terms. 
Such terms are defined in a natural way by the differences between SO(3,2) 
generators and the corresponding generators of the Poincar\'{e} group. 
In a previous article \cite{ws} the author has shown that these  
terms generate interactions between particles, with characteristics of 
the empirical fundamental interactions, including gravitation.

\section{Spinors and spin networks}

Before we come to spin networks, let us consider a large system of, 
say $10^{80}$, binary elements. 
Each of these elements exists in either of two \begin{em}states\end{em} 
that we call `up' and `down'. 
Let $|u\rangle$ and $|d\rangle$ be the symbols for these states.

We can represent these states mathematically in vector form:
\begin{equation}
|u\rangle =  \left( \begin{array}{*{1}{c}} 1 \\ 0 \end{array} \right)  
 \;\; \mbox{ and } \;\;
|d\rangle = \left( \begin{array}{*{1}{c}} 0 \\ 1 \end{array} \right) .  
                                                                \label{2-1}
\end{equation}
(Here the equal-sign means `represented by'.)
The matrix
\begin{equation}
\sigma_3 = \left( \begin{array}{*{2}{c}} 1 & 0 \\ 0 & -1 \end{array} \right)
                                                                \label{2-2}  
\end{equation}                       
can be used to set up an \begin{em}eigenvalue equation\end{em}
\begin{equation}
\sigma_3 |z\rangle = \zeta |z\rangle ,                          \label{2-3}
\end{equation}
where $|z\rangle$ can be either $|u\rangle$ or $|d\rangle$, and $\zeta$
are the corresponding eigenvalues $+1$ and $-1$, respectively.
We can also define the \begin{em}expectation value\end{em} of $\sigma_3$ by
\begin{equation}
\langle z| \sigma_3 |z\rangle = \zeta ,                         \label{2-4}
\end{equation}
where $\langle z|$ stands for the transposed vector of $|z \rangle$.

The matrix
\begin{equation}
\sigma_1 = \left( \begin{array}{*{2}{c}} 0 & 1 \\ 1 & 0 \end{array} \right), 
                                                                \label{2-5}
\end{equation}
applied to $|u\rangle$ or $|d\rangle$, changes $|u\rangle$ into $|d\rangle$ 
and $|d\rangle$ into $|u\rangle$. 
Since neither
$|u\rangle$ nor $|d\rangle$ is preferred, we can say that $\sigma_1$ defines 
a discrete \begin{em}symmetry transformation\end{em} on the states 
$|u\rangle$, $|d\rangle$. 

Let us now extend the definition of a state by allowing also
linear combinations of the form
\begin{equation}
|s\rangle = \alpha \, |u\rangle + \beta \, |d\rangle  ,         \label{2-6}
\end{equation}
where $\alpha$ and $\beta$ are complex numbers with
\begin{equation}
\alpha^* \alpha + \beta^* \beta = 1 .                           \label{2-7}
\end{equation}
We will call such a generalized state a (two-component) 
\begin{em}spinor\end{em}. 
By this step we have created a complex \begin{em}vector space\end{em} with 
$|u\rangle$ and $|d\rangle$ serving as basis vectors. 
We define a scalar product of two states  
\begin{eqnarray}
|s_1\rangle &=& \alpha_1 \, |u\rangle + \beta_1 \, |d\rangle  ,  \\
|s_2\rangle &=& \alpha_2 \, |u\rangle + \beta_2 \, |d\rangle     \label{2-8}
\end{eqnarray}
by 
\begin{equation}
\langle s_1 | s_2 \rangle = \alpha^*_1 \alpha_2 + \beta^*_1 \beta_2  .
                                                                 \label{2-9}
\end{equation}
This extends the vector space to a \begin{em}Hilbert space\end{em}.

By the extension of the description of a binary element, we have
not introduced any new physical degree of freedom.    
Still only the basis states have a physical meaning.
However, the Hilbert space approach allows some mathematical operations that 
are not possible in the binary system. 
We can, for example, ``close" the algebra of the matrices $\sigma_1$ and
$\sigma_3$ with respect to the commutation product and obtain a third matrix 
$\sigma_2$ by means of
\begin{equation}
[\sigma_3, \sigma_1] = i \sigma_2                                \label{2-10}
\end{equation}
in the form
\begin{equation}
\sigma_2 = \left( \begin{array}{*{2}{c}} 0 & -i \\ i & 0 \end{array} \right).
                                                                 \label{2-11}
\end{equation}
It is well-known, that the matrices
$\sigma_1$, $\sigma_2$ and $\sigma_3$, the Pauli-matrices, satisfy the 
commutation relations of SU(2). 
They serve as infinitesimal generators for SU(2) transformations on the 
Hilbert space of two-component spinors.  
SU(2) allows for continuous symmetry transformations of states within 
the Hilbert space, in contrast to only discrete ones between the states of the
original binary elements.
Thus SU(2) extends the basic discrete symmetry into the ``interpolating" 
Hilbert space.
It is, therefore, the \begin{em}symmetry group\end{em} of this Hilbert
space.
The infinitesimal generators of SU(2) are considered as 
\begin{em}observables\end{em} in the quantum mechanical sense.

This step has laid the basis for a quantum mechanical description 
of binary elements. 
The introduction of a Hilbert space does not change any of the ``physical" 
properties of binary elements. 
It merely offers a new way of description, similar to the introduction of, 
for example, a potential in classical mechanics.

A \begin{em}system\end{em} of spinors is then quantum mechanically described 
by a Hilbert space that is given by the (antisymmetrized) direct product of  
individual spinor states.
A set of identical elements that are defined in this way and allow for 
the representation of a certain Lie group constitute a
\begin{em}spin network\end{em} \cite{jcb}. 
Our SU(2) based system can be considered as the simplest type of a spin 
network.

Within such a spin network we can define sub-networks by assembling a large
number of spinors. Within their Hilbert spaces we can select states with large
SU(2) quantum numbers. Large quantum numbers allow for a quasi-continuous 
orientation of states with respect to the ``direction" defined by the state 
of a given single spinor.

Assume that we rotate a sub-network relative to a spinor by an angle of 
$\pi/2$.
Then by applying the corresponding SU(2) transformation to the state of the
spinor we can map this rotation into the Hilbert space of the spinor.
Without knowing details of the measuring process, we can conclude from 
symmetry arguments, that the outcome of a subsequent measurement of the spinor
state \begin{em}relative\end{em} to the orientation of the sub-network is 
non-predictable. 
Of course, this is familiar from the basics of quantum mechanics. 
This means that we have a 50:50 per cent chance to measure either an ``up" or
a ``down" state. 
Therefore, the coefficients of a spinor state (\ref{2-6}) carry information
about the relation between spinor and an observer in the sense of probability 
amplitudes. 

Coming back to Bohr's principle, we can say that a description of
a single spinor \begin{em}relative\end{em} to a macroscopic sub-network, has 
become possible by the extension of the basic states to a vector space. 
Or if we formulate this statement the other way round: if we want to
represent a non-discrete symmetry on a discrete system, a suitable way to 
achieve this is to turn to a quantum mechanical description of the system. 
We will come back to this point in section VII.

\section{From spinors to de~Sitter group}

In the last section we have taken for granted that there is only one way
to code the information ``up" and ``down" as given by (\ref{2-1}).
However, we could have defined just as well 
\begin{equation}
|u\rangle =  \left( \begin{array}{*{1}{c}} 0 \\ 1 \end{array} \right)  
 \; \mbox{ and } \;                                             \label{3-1}
|d\rangle = \left( \begin{array}{*{1}{c}} 1 \\ 0 \end{array} \right) .                                                          
\end{equation}

It is a common experience in software development that programmers
fail to agree on the way of how they code e.g. the outcome of a software 
procedure.
Some may return $-1$ in the error case, other $+1$. 
Therefore, if we count errors, we cannot simply look for an outcome of 
$-1$, but also have to consider the other way of coding.
Of course, a good project management should prevent such situations arising.

Fortunately, the codings (\ref{2-1}) and (\ref{3-1}) represent the only ways
of coding a piece of binary information in a binary system. 
For a single binary element the way of coding is not relevant.
However, if we want to compare the information contained in several
elements of a binary system, we either have to agree on one unique way
of coding or we have to provide information about the way of coding with
each binary element.

In the first case we can use either (\ref{2-1}) \begin{em}or\end{em}
(\ref{3-1}).
In the second case the description of the information contained in a
single element requires a four-component state:
\begin{eqnarray}
|u_1\rangle &=& \left(\begin{array}{*{1}{c}} 1 \\ 0 \\ 0 \\ 0 \end{array}\right)  
 \; \mbox{ , } \;
|d_1\rangle  =  \left(\begin{array}{*{1}{c}} 0 \\ 1 \\ 0 \\ 0 \end{array}\right)  
\label{3-2a}\\
\nonumber \\
|u_2\rangle &=& \left(\begin{array}{*{1}{c}} 0 \\ 0 \\ 0 \\ 1 \end{array}\right)  
 \; \mbox{ , } \;
|d_2\rangle  =  \left(\begin{array}{*{1}{c}} 0 \\ 0 \\ 1 \\ 0 \end{array}\right).  
                                                                \label{3-2b}
\end{eqnarray}
These four-component objects can be understood as obtained from a direct sum of 
the two alternative codings of the same type of binary information.
In the following we will call these objects \begin{em}Dirac spinors\end{em}.

The second case has to be considered the general one, since it does not
presume an agreement on the way of coding. 
Such an agreement, in a global sense, would be hardly compatible with 
relativistic causality.
In other words, we cannot expect a global ``project management" as far as
the coding is concerned.
Therefore, we consider the second case as the realistic one.

The following is based on four-component states as defined by (\ref{3-2a})
and (\ref{3-2b}).

Again we can form a Hilbert space by interpolation between the basic states.
Its symmetry group is considerably larger than in the case
of two-component spinors.
Instead of Pauli matrices we now have $4\times4$ spin matrices: 
\begin{equation}
\sigma_{ij} = \epsilon_{ijk} \left(\begin{array}{*{2}{c}} 
\sigma_k & 0 \\ 0 & \sigma_k \end{array}\right),
\; i,j,k = 1,2,3,                                               \label{3-3}                                                                
\end{equation}
where $\epsilon_{ijk}$ is the permutation symbol.
The matrix
\begin{equation}
\gamma^0 = 
\left( \begin{array}{*{2}{c}} I & 0 \\ 0 & -I \end{array} \right), 
                                                                \label{3-4}
\end{equation}
where $I$ is the $2\times 2$ unit matrix,
delivers an eigenvalue of $+1$, if applied to the first group of spinors 
(\ref{3-2a}), and $-1$, if applied to the second (\ref{3-2b}).
By analogy with $\sigma_1$ we can define a matrix
\begin{equation}
\tau = \left( \begin{array}{*{2}{c}} 0 & I \\ I & 0 \end{array} \right), 
                                                                \label{3-4a}
\end{equation}
which interchanges the first and second group of spinors.
Since none of the codings is preferred, the interchange operation again 
defines a symmetry transformation.
Then also the operation of combined application of $\sigma_1$ and $\tau$ 
is a symmetry operation. 
It is given by the matrix
\begin{equation}
\gamma^1 = 
\left( \begin{array}{*{2}{c}} 0 & \sigma_1 \\ \sigma_1 & 0 \end{array}\right).
\label{3-5} 
\end{equation}

When we close the algebra of the matrices $\gamma$ and $\sigma$ defined so far, 
with respect to the commutation product, we find additional matrices
\begin{equation}
\gamma^k = 
\left( \begin{array}{*{2}{c}} 0 & \sigma_k \\ \sigma_k & 0 \end{array}\right) 
                                                                \label{3-6}
\end{equation}
and
\begin{equation}
\sigma^{0k} = -\sigma^{k0} = \left( \begin{array}{*{2}{c}} 0 & i\sigma_k \\
       -i\sigma_k & 0 \end{array}\right).        \label{3-6a}
\end{equation}
We can combine the indices $0$ and $k$ to an index $\mu = 0,\ldots,3$, and
use $g_{\mu\nu}=$ diag $(+1,-1,-1,-1)$ in the usual way to raise and lower
indices.

The matrices (\ref{3-4}) and (\ref{3-6}) are Dirac's $\gamma$-matrices in the 
so-called standard or Dirac representation. They satisfy the well-known 
anti-commutation relations
\begin{equation}
\{\gamma_\mu, \gamma_\nu \} = 2 g_{\mu\nu} .                    \label{3-7a}
\end{equation} 
and the commutation relations
\begin{equation}
\frac{i}{2} \, [\gamma_\mu, \gamma_\nu ] = \sigma_{\mu\nu} .    \label{3-7b}
\end{equation} 

A scalar product can be defined, as known from the treatment of the Dirac 
equation,
\begin{equation}
\langle \bar{a} | b \rangle  \mbox{   with   } 
\langle \bar{a} | = \langle a | \gamma^0  .                     \label{3-8}
\end{equation}

The $4\times4$-matrices
$s_{\mu\nu}$ and $s_{\mu4}$, built from Dirac matrices, 
\begin{equation}
s_{\mu\nu} :=\, \case{1}{2} \sigma_{\mu\nu}             
\mbox{   and  }  
s_{\mu4} :=\, \case{1}{2} \gamma_\mu                            \label{3-9}
\end{equation} 
form an irreducible representation of SO(3,2). 
The proof is by verifying the commutation relations of SO(3,2):
\begin{equation}
[s_{\mu\nu}, s_{\rho\sigma}] = 
-i[g_{\mu\rho} s_{\nu\sigma} - g_{\mu\sigma} 
s_{\nu\rho} + g_{\nu\sigma} s_{\mu\rho} 
- g_{\nu\rho} s_{\mu\sigma}] \mbox{ , }                         \label{3-10}
\end{equation}
\begin{equation}
[s_{\mu4}, s_{\nu4}] = -i s_{\mu\nu} \mbox{ , }                 \label{3-11}
\end{equation}
\begin{equation}
[s_{\mu\nu}, s_{\rho4}] 
= i[g_{\nu\rho} s_{\mu4} - g_{\mu\rho} s_{\nu4}]  .             \label{3-12}	 
\end{equation} 

The transformations of SO(3,2) extend the discrete symmetry operations
of the general binary system, to the interpolating Hilbert space of
Dirac spinors.
Therefore, SO(3,2) is the basic symmetry group of Dirac spinors.
A subgroup of SO(3,2) is the homogeneous Lorentz group SO(3,1) with the
commutation relations (\ref{3-10}).
The common subgroup SO(3) reflects the SU(2) symmetry of the two-component 
spinor parts of the Dirac spinor.

A \begin{em}system\end{em} of Dirac spinors is described by the direct product 
of individual Hilbert spaces. 
The operators  
\begin{equation}
S_{\mu 4} = \sum s_{\mu 4},  \; \;  S_{\mu\nu} = \sum s_{\mu\nu} 
                                                                \label{3-13}
\end{equation}
then generate SO(3,2) symmetry transformations and are considered as 
observables.

\section{Poincar\'{e} group and space-time}

Consider now a \begin{em}large\end{em} sub-network within a spin network of 
Dirac spinors. 
From these spinors we can construct states with large quantum numbers.
These quantum numbers can then be regarded as quasi-continuous. 
Assume that we have constructed these states in such a way that
the expectation values of the operators $S_{\mu 4}$ are large 
compared to those of $S_{\mu\nu}$.

Then we can apply group contraction \cite{iw,fg} to this sector of the 
Hilbert space, and thus obtain the Poincar\'{e} group P(3,1) as an approximate 
symmetry group.
The operators $S_{\mu 4}$ are then approximated by translation operators
\begin{equation}
S_{\mu 4} \rightarrow P_\mu                                     \label{4-1}
\end{equation}
with
\begin{equation}
[P_\mu, P_\nu] = 0                                              \label{4-2}
\end{equation}
and
\begin{equation}
[S_{\mu\nu}, P_\rho] 
= i[g_{\nu\rho} P_\mu - g_{\mu\rho} P_\nu]  .                   \label{4-3}	 
\end{equation} 
The quasi-continuous spectrum of quantum numbers is now replaced by the
continuous one of the operators $P_\mu$.

Eigenstates of $P_\mu$ can be used to form new states $|X\rangle$ that are
\begin{em}localized\end{em} in spacelike directions
\begin{equation}
|X\rangle = (2\pi)^{-3/2} \int\limits^\infty_{-\infty}\! d^3P 
\: e^{i x^\mu P_\mu}\, |P\rangle.                                \label{4-4}
\end{equation}
The operators $P_\mu$ generate translations when applied to the states 
$|X\rangle$
\begin{equation}
e^{ia^\mu P_\mu} \, |X\rangle = |X + a\rangle .                  \label{4-5}
\end{equation}
The parameter space of the offsets $a$ then defines a 4-dimensional 
\begin{em}space-time\end{em} continuum with Minkowskian metric. 

This means that we have derived space-time as a property of large 
sub-networks in the form of a tangential space-time continuum.
Tangential means: it is a flat-space approximation, obtained by group 
contraction, valid for the neighbourhood of a given point in space-time. 
In section VI we will indicate how more accurate approximations can be 
obtained.

\section{Spinors in space-time}

After having introduced the concept of space-time for large or
``macroscopic" sub-networks, Bohr's principle motivates us to try to find a 
description of a single spinor relative to the positions of sub-networks. 

In the following, we will use the term ``macrosystem" for a macroscopic 
sub-network that is described by the Poincar\'{e} invariant approximation.

Let $|P\rangle$ be the state of a macrosystem with momentum $P$, 
and let $|s\rangle$ be the state of a separate Dirac spinor that is not 
part of the macrosystem. 
Then the combined state is obtained from the direct product 
$|P\rangle |s\rangle$. 
For sake of clarity we add the operators corresponding to $S_{\mu4}$
\begin{equation}
(P_\mu + s_{\mu4}) |P\rangle |s\rangle   .                      \label{5-1}
\end{equation}
Since the Dirac spinor can also be considered as part of an
extended macrosystem, there exists an alternative representation in 
the form 
\begin{equation}
P_\mu |P'\rangle = P_\mu |P + p_s\rangle ,                      \label{5-2}                  
\end{equation}   
which describes the macrosystem including the spinor. 
In the Poincar\'{e} invariant approximation, 
on which we have based this consideration,
$P'$ can differ from $P$ only by an incremental 4-momentum $p_s$, which 
may depend on the ``internal" quantum numbers of $|s\rangle$.
The momentum $p_s$ can be considered as the effect on the macrosystem that is 
caused by the inclusion of the spinor.

Let us now decide that in the following the momentum operator $P_\mu$ shall 
act only on the original macrosystem (as in (\ref{5-1})). To compensate for
this decision we introduce a second momentum operator $p_\mu$ that acts on 
the momentum $p_s$.
Suppose that the momentum state on the rhs of (\ref{5-2}) is given in the form
\begin{equation}
e^{ix_\mu(P_M^\mu+p^\mu_s)} = e^{ix_\mu P_M^\mu} \;\; e^{ix_\mu p^\mu_s}, 
\label{5-3}
\end{equation}
where $P_M^\mu$ and $p^\mu_s$ are quantum numbers of macrosystem and spinor,
respectively.
Then we can split (\ref{5-2}) in the following way 
\begin{equation}
(P_\mu + p_\mu) |P\rangle |p_s\rangle .                         \label{5-4}                  
\end{equation}

When we compare (\ref{5-1}) with (\ref{5-4}) we find the
following correspondence 
\begin{equation}
s_{\mu4} |s\rangle \Longleftrightarrow p_\mu |p_s\rangle .      \label{5-5}                  
\end{equation}
We can understand this in the following way:
In combination with the state of a macrosystem,
the action of $s_{\mu4}$ on $|s\rangle$ is simulated by the action of $p_\mu$
on an auxiliary state $|p_s\rangle$.

The dependency on $s$ makes $|p_s\rangle$ again a four-component state. 
Therefore, we can map this state into the Hilbert space of Dirac spinors. 
Let $|\psi_s(p)\rangle$ be the image of $|p_s\rangle$.
Then $|\psi_s(p)\rangle$ is a Dirac spinor defined as a superposition
of the four basic states with coefficients depending on $p_s$.
This mapping has to be covariant with respect to transformations of the 
Poincar\'{e} group.
Therefore, to determine the coefficients, we need a rule that is invariant 
with respect to P(3,1). 
The only invariant rule that can be formed from the characteristic operators 
$p_\mu$ and $s_{\mu4}$ of relation (\ref{5-5}) and that is linear in both 
operators can be written, with $s_{\mu4} = \frac{1}{2}\gamma_\mu$, as
\begin{equation}
(\gamma^\mu p_\mu - m) |\psi_s(p)\rangle = 0 .                   \label{5-8}            
\end{equation}
This is the Dirac equation with a mass of m.
So the Dirac equation appears here as a relation that links representation
(\ref{5-4}) to (\ref{5-1}).

A value for the mass $m$ is obtained by the following consideration.
The operator $p^\mu p_\mu$ ``simulates" the operator 
$\frac{1}{4}\gamma^\mu \gamma_\mu$, which evaluates to the unit matrix.
Therefore, a reasonable estimate for $m$ is the value of 1 (measured in units
of quantum numbers). 

By this manipulation we have formally given a single spinor a momentum in 
relation to an arbitrary macrosystem. 
This does not mean that the spinor has obtained additional degrees of freedom.
There are still only four independent states in the Hilbert space of a spinor.
The momentum information is instead rather coded into the coefficients of the 
spinor states.
These coefficients are essentially phase factors and reflect the way how the 
spinor is coupled to a macrosystem.

Because we have ``borrowed" the momentum $p_s$ from the macrosystem, the 
transformation properties of $p_s$ with respect to the Poincar\'{e} 
group are identical to that of $P$.
From the momentum states $|\psi_s(p)\rangle$ we again can form states that are 
formally localized in spacelike directions. 
Then the translations of the Poincar\'{e} group will move these localized 
states in space-time. 
Such a state may ``start" from a macrosystem A, and then ``arrive" 
at the position of another macrosystem B. Again we can form a direct product
with macrosystem B, and obtain the effect which this spinor causes to B. 
However, between A and B the position of the spinor is not defined. 
This does not cause any problems, but reflects the fact that we do not have 
information about the position of the spinor between A and B.
But let us insert an experimental setup C in between A and B, to observe 
the spinor on its way from A to B. Then we can form a product state of
the spinor and the setup C, and verify that it was in fact on its way from 
A to B, but was caught at C.

Thus we have arrived at a logically consistent description of a 
\begin{em}spinor in space-time\end{em}. 
The apparent space-time properties of the spinor do not reflect 
degrees of freedom of the spinor itself, but stand for the 
\begin{em}relation\end{em} of the spinor to macrosystems. 
If there where no other spinors than the one under consideration, 
we would not have been able to introduce any space-time properties.

\section{Interactions}

The flat-space approximation obtained by group contraction cannot be
the final answer to the kinematical properties of a SO(3,2) based
spin network. The following considerations will show why.

Let us come back to a macroscopic sub-network, described by momentum
$P$, and a single spinor $s$, which form a common state $|P,s\rangle$. 
In flat-space approximation, translations of this state are generated by 
the momentum operator $P_\mu$.
In an exact SO(3,2) description, we have instead a state $|S,s\rangle$, 
and the operator 
\begin{equation}
S_{\mu4} + s_{\mu4}                                             \label{6-1}
\end{equation}
generates symmetry transformations that correspond to the translations in 
flat-space approximation.
In performing the contraction limit, where the contributions of $S_{\mu4}$
are assumed to grow over all limits, the contributions of $s_{\mu4}$ to the 
translations usually are ignored relative to those of $S_{\mu4}$.

But in the last section the situation is different.
Although the momentum operator $p_\mu$ has been derived from the momentum 
$P_\mu$ of the macroscopic system, the contribution of $p_\mu$ can no longer
be considered large with respect to that of $s_{\mu4}$. 
In fact, since
$p_\mu$ stands for the effect of $s_{\mu4}$ on the macrosystem, both
operators have the same magnitude.
This requires an improvement of the approach in the last section by taking 
into account SO(3,2) specific elements.

We may start from SO(3,2) states $|S\rangle |s\rangle$ that 
are given by the direct product of a quasi-continuous representation 
of a macroscopic sub-network and a spinor representation.
By reduction with respect to SO(3,2) we obtain
\begin{equation}
|S\rangle |s\rangle \rightarrow \sum_s |S_s,s\rangle             \label{6-2}
= \sum_s |S + d_s, s\rangle .
\end{equation}
The target is then to factorize the state on rhs of (\ref{6-2}) such that
\begin{equation}
\sum_s |S + d_s, s\rangle = |S\rangle \, \sum_s |d_s,s\rangle . \label{6-3}
\end{equation}
Then again the sub-network $|S\rangle$ can be separated and we obtain a 
SO(3,2) description of a spinor relative to a sub-network by states
\begin{equation}
|d_s,s\rangle . \label{6-4}
\end{equation}
This description has the form of a direct product of a quasi-continuous 
representation and a spinor representation.
Whether or not this target can generally be reached is not known.

However, an approach starting from exact SO(3,2) states, as just sketched, 
is not what is really required, for the following reason:
Experimental setups are macrosystems in a Poincar\'{e} invariant environment.
If we try to apply corrections to the kinematics which are required by SO(3,2) 
symmetry, we, nevertheless, have to measure the consequences of these 
corrections \begin{em}relative to a Poincar\'{e} invariant basis\end{em}. 
Therefore, we need a theoretical approach that is adaptive to this situation.

This demand can be satisfied by a perturbational approach that treats the 
differences between the generators of SO(3,2) and P(3,1), for example the term 
$s_{\mu4}$, as perturbations to a Poincar\'{e} invariant basic system.
Whether or not the perturbation algorithm converges to separable states
in the sense of (\ref{6-3}) is of secondary interest.
Of physical interest are rather, the consequences of corrections caused by 
SO(3,2) symmetry. 
These corrections are obtained by employing SO(3,2) operators, but do not 
require a knowledge of the exact SO(3,2) states.

Such a perturbative approach has been described in a recent article by the 
author \cite{ws}.
The article is based on a spin network with SO(3,2) representations in 
the sense of (\ref{6-4}). It treats the spin network by perturbation
methods relative to a Poincar\'{e} invariant basis.
This leads to the identification of four types of perturbation terms. 
It is shown that these terms give rise to interactions between particles 
with characteristics of the four empirical interactions: electromagnetic, 
weak, chromodynamic and gravitational.

\section{Bohr's principle and quantum mechanics}

We have intentionally started from an set of binary elements rather 
than from a spin network. We then have introduced a Hilbert space 
that interpolates between the states of the binary elements,
and have obtained SU(2) and SO(3,2) based spin networks.

Macroscopic sub-networks are described by quasi-continuous representations 
and allow, for example, quasi-continuous rotations in space. 
According to Bohr's principle, we had to find a way to describe the
discrete system of a single spinor \begin{em}relative to a macroscopic\end{em}
setup that can be randomly rotated in space.
Obviously, a practicable way is to use a description that interpolates
between the basic states of the microscopic system.

So our answer to the question ``Why quantum mechanics?" is simply this: 
Quantum mechanics is a way to fulfil Bohr's principle in the microscopic 
domain.

We have introduced quantum mechanics at the binary level in a simple and 
rather formal way.
If we had not known quantum mechanics before, we could have hardly anticipated
the extreme power of quantum mechanics in general. 
Under this perspective, the success of quantum mechanics can be ascribed 
mainly to the following properties:
Firstly, the quantum mechanical calculus can be seamlessly extended from 
binary systems to complex composed systems, by forming direct products of 
Hilbert spaces of subsystems.
Secondly, quantum mechanics keeps track of probabilities, and 
composed systems inherit this property from their parts.
This delivers information of statistical nature in situations 
where no deterministic information is available. 
Lastly, in ``classical" limits probabilistic properties become 
deterministic and may even manifest themselves in form of additional 
independent degrees of freedom.
An example in this context are momentum and position.
They become independent degrees of freedom, although position, at the quantum 
mechanical level, is defined only by superposition of momentum states.

So the introduction of an interpolating Hilbert space appears as a
mathematical trick that in the end is more than justified by practical 
results.

\section{Conclusion}

Bohr's principle has guided us from a most general binary structure, to 
physical concepts that in the end, allows us to describe observations 
relative to experimental setups. 
These concepts include the basics of quantum mechanics, space-time, 
particles and interaction. 

It is very satisfactory, that as a consequence of this approach a Minkowskian 
space-time continuum is obtained. 
Thus space-time appears as a derived phenomenon, that stands for a way 
of describing subsystems within binary systems relative to each other.

Note that we have obtained space-time as a property of 
\begin{em}macroscopic\end{em} systems. 
This is in contrast to (so far unsuccessful) attempts to derive space-time
from discrete structures at Planck scales.
Nevertheless, our approach is also basically a discrete one,
but discrete at microscopic scales rather than at Planck scales.
 
The symmetry operations that have been made available by the basic symmetry
group SO(3,2) deliver the basis for rearrangements of relations and especially
for different reference frames. 
Thus Bohr's principle of relativity can be considered as the root of 
relativistic principles, including special and general relativity,
and also of dynamical processes within spin networks. 

But first of all Bohr's principle has guided us to a description of 
\begin{em}microscopic phenomena relative to macroscopic setups\end{em} 
by interpolating Hilbert spaces. 
Therefore, quantum mechanics can be understood as a necessary logical 
result of the application of Bohr's principle to the microscopic domain.

Our results exhibit a general binary structure as a basis, from which 
essential aspects of particle physics can be derived.
Is then a binary system the ultimate answer to our search for basic 
structures in nature?
Does it describe some kind of a pre-physical aspect of nature, or is it rather
a logical consequence of our continuous search for more and more elementary 
structures?
We know from information technology that we arrive at binary structures, 
when we divide information into smaller and smaller pieces. 
Do we come across the same situation in particle physics?

To quote another dictum of Bohr's: "Physics is not about reality, 
but about what we can say about reality".
In this sense, binary systems can be regarded as a general means to represent
our knowledge about nature at the most elementary level. 
But then the binary system itself is not part of the knowledge. 
It is merely an unspecific logical scheme, comparable to the binary system in
computer technology, that is used to represent information but is not part of 
the information. 
Since binary information cannot be divided further, there cannot be any 
knowledge beyond what can be coded into the binary structure.

Therefore, we cannot say that nature has the structure of a binary system. 
The binary system rather reflects our way of doing science. 
We typically practise science by dividing knowledge into smaller and
smaller parts.
At least we can say that our knowledge about nature at microscopic 
scales can be represented within a discrete, binary logical scheme.
At the same time we have to accept the fact that the binary structure sets a 
logical limit to further refining of knowledge.
Also: there are no ``laws of nature" at the binary level other than the laws 
of mathematical logic. 

If this is the ``final answer", then quantum mechanics, space-time, particles
and interactions are inevitable logical consequences, not of fundamental laws
of nature, but of the way in which we do physics. This way is determined
by two of our behaviour patterns:
firstly, by our habit to divide information into smaller and smaller parts,
and, secondly, by the application of Bohr's principle of relativity to observe 
and describe relations between these parts.

This view illustrates another famous dictum of Bohr's: 
``It is the task of science to reduce deep truths to trivialities".

\renewcommand{\baselinestretch}{1.1}

\end{document}